\documentclass{epl}

\usepackage{bm}
\usepackage{cite}
\usepackage{amsmath}
\usepackage{graphicx}

\title{On the pearl size of hydrophobic polyelectrolytes}
\shorttitle{On the pearl size...}
\author{D. Baigl\inst{1} \and M. Sferrazza\inst{1,2} \and C. E. Williams\inst{1}\thanks{E-mail: \email{claudine.williams@college-de-france.fr}}}
\institute{
  \inst{1} Laboratoire des Fluides Organis\'es,
CNRS UMR 7125,
Coll\`ege de France - 11, place Marcelin Berthelot, 75005 Paris, France.\\
  \inst{2} Department of Physics, University of Surrey - Guildford, Surrey GU2 7XH, UK. }

\pacs{82.35.Rs}{Polyelectrolytes} \pacs{81.07.Nb}{Molecular
nanostructures} \pacs{68.47.Mn}{Polymer surfaces}

\begin{document}

\maketitle

\begin{abstract}
Hydrophobic polyelectrolytes have been predicted to adopt an
unique pearl-necklace conformation in aqueous solvents. We present
in this Letter an attempt to characterise quantitatively this
conformation with a focus on $D_p$, the pearl size. For this
purpose polystyrenesulfonate (PSS) of various effective charge
fractions $f_{eff}$ and chain lengths $N$ has been adsorbed onto
oppositely charged surfaces immersed in water in condition where
the bulk structure is expected to persist in the adsorbed state.
\emph{In situ} ellipsometry has provided an apparent thickness
$h_{app}$ of the PSS layer. In the presence of added salts, we
have found: $h_{app}\sim aN^{0}f_{eff}^{-2/3}$ ($a$ is the monomer
size) in agreement with the scaling predictions for $D_p$ in the
pearl-necklace model if one interprets $h_{app}$ as a measure of
the pearl size. At the lowest charge fractions we have found
$h_{app}\sim aN^{1/3}$ for the shorter chains, in agreement with a
necklace/globule transition.
\end{abstract}

\section{Introduction}
Polyelectrolytes are called hydrophobic when water is a poor
solvent for the backbone. Because many artificial or natural
macromolecules may have some intrinsic hydrophobicity,
understanding the behaviour of hydrophobic polyelectrolytes
constitutes a challenging area for fundamental and practical
physical studies \cite{CW_houches}. The most spectacular property
of hydrophobic polyelectrolytes is the pearl-necklace conformation
predicted for the single chain. This unusual conformation was
obtained theoretically by Dobrynin \emph{et al.}
\cite{Dobrynin_1996,Dobrynin_1999} by analogy with the Rayleigh
instability of a charged droplet \cite{Rayleigh_1882,Kantor}.
Indeed a neutral hydrophobic polymer when immersed in water forms
a spherical globule (like a drop of oil in water) in order to
minimise the interfacial area with the solvent. When charges are
added to the polymer the associated electrostatic energy of the
globule increases until it balances the interfacial energy at some
critical charge fraction $f_C$ . Beyond $f_C$ the globule splits
into two smaller ones thus increasing the interfacial area but
decreasing the electrostatic energy. Contrary to the case of a
charged droplet, the presence of connected monomers hinders
infinite separation and an elongated string is formed between the
pearls. Therefore the resulting structure is a water-soluble
pearl-necklace composed of dense pearls connected by narrow
strings (most of the monomers are contained in the pearls).
Scaling theory provides the description of hydrophobic
polyelectrolytes properties \cite{Dobrynin_1996,Dobrynin_1999}
which we summarise here. The pearl density $\rho$ is that of a
globule of a neutral chain in a poor solvent $\rho \sim
(a^{2}\xi_{\rm T})^{-1}$ where $a$ is the monomer size and
$\xi_{\rm T}$ is the thermal blob size.  A pearl has a charge of
$\rho D_{p}^{3} f_{eff} \rm{e}$ where $f_{eff}$ is the effective
linear charge density ($f_{eff}=0.1$ or $10 \% $ means that every
tenth monomer is effectively charged in average) and e is the
elementary charge. When the electrostatic energy balances the
interfacial energy (proportional to $D_p^2$), we have:
\begin{equation}
\frac{(\rho D_{p}^{3}f_{eff} {\rm e})^{2}}{\epsilon D_{p}} \sim
k_{\rm{B}}T \frac{D_{p}^{2}}{\xi_{\rm{T}}^{2}} \label{pearlsize}
\end{equation}
and the pearl size scales as:
\begin{equation}
D_{p} \sim a (\frac{l_{\rm B}}{a})^{-1/3} f_{eff}^{-2/3}
\label{pearl scaling}
\end{equation}
$l_{\rm B}$ is the Bjerrum length ($\sim {\rm e}^2/(\epsilon
k_{\rm B}T)$) and equals 0.7 nm in pure water at $25
\un{^{\circ}C}$. The distance between pearls, the string length
$l_{string}$, results from the balance of repulsion between two
neighbouring pearls with surface energy of the string,
proportional to $l_{string}$:
\begin{equation}
\frac{(\rho D_{p}^{3}f_{eff} {\rm e})^{2}}{\epsilon l_{string}}
\sim k_{\rm B}T \frac{l_{string}}{\xi_{\rm T}} \label{lstring1}
\end{equation}
\begin{equation}
l_{string}\sim a (\frac{l_{\rm B}\xi_{\rm T}}{a^2})^{-1/2}
f_{eff}^{-1}\label{lstring2}
\end{equation}

Simulations also showed a pearl-necklace conformation
\cite{Micka_1999} and more recently confirmed that pearl formation
was due to an instability since pearls were observed to fluctuate
along the chain in number and in size\cite{Limbach_2002}.
Experimentally, Essafi \emph{et al.} studied
poly(styrene-\emph{co}-sodium styrenesulfonate), abbreviated PSS,
as a model hydrophobic polyelectrolyte. By measuring the
fluorescence emission of pyrene probes they observed the existence
of low dielectric constant nano-regions dispersed along the chains
\cite{Essafi_1995} in agreement with the presence of hydrophobic
pearls on the chains. They measured also the effective charge
fraction as a function of the chemical charge fraction by
osmometry \cite{Essafi_thesis}. In a range of charge fraction
where hydrophilic polyelectrolytes followed rather well the
so-called Manning-Oosawa condensation theory
\cite{Manning_1969,Oosawa_1971}, they found for PSS a further
reduction of $f_{eff}$ according to the empirical linear law:
\begin{equation}
f_{eff} (\%)= \frac{35(f(\%)-27)}{73} \label{feffvsf}
\end{equation}
This effect is compatible with the pearl-necklace model  if one
also takes into account ionomer effects due to the low local
dielectric constant in the pearls. Further experimental evidences
for the pearl-necklace conformation may be found
in~\cite{CW_houches} and references within.

However quantitative investigations of the single chain
pearl-necklace conformation are difficult to perform
experimentally on chains in solution. This is due in part to the
low contrast in scattering experiments which prevents dilute
solution measurement and also to the fluctuating nature of
polyelectrolyte solutions. An interesting alternative consists in
looking at chain conformation in the adsorbed state. Th\'{e}odoly
\emph{et al.} showed that hydrophobic polyelectrolytes adsorbed
spontaneously at a hydrophobic interface but adopted a flat
conformation whatever charge fraction, the pearls unfolding on the
surface. The film thickness was then independent of
$f$~\cite{Theodoly_2001}. For the investigation reported here
hydrophobic polyelectrolytes have been adsorbed electrostatically
onto oppositely charged solid surfaces in experimental conditions
where the pearl-necklace conformation is expected to persist on
the surface \cite{Dobrynin_2002}. The thickness of the adsorbed
monolayer has been measured by \emph{in situ} ellipsometry and
interpreted afterwards in terms of the pearl size $D_{p}$.

In this study we used also PSS as a model hydrophobic
polyelectrolyte. Control of the chemistry \cite{Baigl_2002}
allowed us to synthesise and characterise a series of well defined
monodisperse ($M_W/M_N<1.3$) hydrophobic polyelectrolytes of
various chain lengths and charge fractions (linear charge
densities) as summarised in Table \ref{tablePSS}. Fully charged
PSS's ($f=100\%$) were purchased from Fluka.

\begin{table}
\caption{PSS samples of various chain lengths $N$, chemical charge
fractions $f$ and effective charge fractions $f_{eff}$. $f_{eff}$
is calculated from $f$ according to the empirical renormalisation
law (eq. \ref{feffvsf})} \label{tablePSS}
\begin{center}
\begin{largetabular}{lllcccrrr}
    $N$   &$f(\%)$ &$f_{eff}(\%)$ &$N$ &$f(\%)$ &$f_{eff}(\%)$ &$N$ &$f(\%)$ &$f_{eff}(\%)$
\\  120     &34     &3.4        &930     &33     &2.9       &1320   &36 &4.3
\\  120     &54     &13         &930     &41     &6.7       &1320   &53 &12
\\  120     &64     &18         &930     &46     &9.1       &1320   &71 &21
\\  120     &94     &32         &930     &53     &12        &1320   &91 &31
\\  \cline{1-3}
    410     &39     &5.8        &930     &64     &18        &1320   &100 &35
\\  \cline{7-9}
    410     &56     &14         &930     &83     &27        &2520   &37 &4.8
\\  410     &71     &21         &        &       &          &2520   &54 &13
\\  410     &100    &35         &        &       &          &2520   &89 &30

\end{largetabular}
\end{center}
\end{table}

PSS solutions were prepared at a polymer concentration of 0.01
mol/L in deionised water (Millipore water) with added salt (NaCl)
or in salt-free conditions. Positively charged surfaces were
prepared by grafting on silicon wafers a self assembled monolayer
(SAM) with terminal amino groups composing the upper surface and
obtained by the reaction of (3-Aminopropyl)-trimethoxysilane on
the silica covering the wafer and previously activated. Silica
activation consisted of 45 minutes $\rm UV-O_{3}$ treatment prior
to 15 minutes exposure under a $\chem{O_2}+\chem{H_2O}$ gas flow.
Silanation was made in liquid phase (toluene). The SAM quality
(thickness $\simeq 1.1$ nm and roughness of the bare silicon) was
controlled in air by ellipsometry and  X-ray reflectivity.
\emph{In situ} ellipsometry measurements were performed using a
variable angle spectroscopic ellipsometer (VASE Woollam Inc., USA)
with a rotating analyzer configuration. The sample was contained
in a liquid cell with thin glass windows perpendicular to the
incident light beam at $72^{\circ}$ with respect to the sample
normal. The ellipsometric angles $\Psi$ and $\Delta$ were measured
as a function of the wavelength $\lambda$ ($400 \un{nm}\leq
\lambda \leq 700 \un{nm}$). A three layer ($\chem{SiO_2}$/SAM/PSS)
optical model was used for the fitting of the ellipsometric
angles. The  thicknesses of the first two layers were kept to the
previously measured values: $\chem{SiO_2}$ in air prior to SAM
grafting, SAM in water before PSS adsorption. The PSS layer has
been modelled
 with an uniform refractive index $n_{\rm PSSlayer}$ and an apparent thickness
$h_{app}$.

First of all it was found that, whatever chain length, charge
fraction or ionic strength conditions, the adsorption of PSS onto
oppositely charged surfaces presented some general features: the
adsorption was very fast and equilibrium was reached after a few
minutes; at equilibrium several consecutive measurements were
performed showing that the layer was stable; when the adsorbing
solution was flushed by a NaCl solution at the same ionic strength
(or pure water in case of salt-free adsorption) the layer
thickness remained constant; when this NaCl solution was flushed
by deionised water there was no measurable change. This shows the
strongly irreversible character of this electrostatically driven
adsorption. In general the layer characteristics only depended on
the bulk conditions (ionic strength) \emph{during} the adsorption
process.

First, PSS's were adsorbed from salted solutions (${\rm
[NaCl]}=0.1$ mol/L). The ionic strength was chosen so that the
Debye length $\lambda_{D}$ was so small ($\lambda_{D} < 1 \rm nm$)
that pearls could come very close without any significant
repulsion but large enough to allow PSS adsorption. If, as theory
predicts, the bulk conformation remains in the adsorbed state, the
adsorbed layer could be imagined as a dense carpet of pearls.

The ellipsometric determination of $h_{app}$ requires the
knowledge of the refractive index of the adsorbed layer
$n_{layer}$. Fits of the ellipsometric angles gave for all samples
a value of $n_{layer}\simeq1.45$. This value is compatible with
that estimated for a compact layer of pearls (remember that the
majority of PSS monomers belong to the dense pearls). Using a
linear approximation $n_{layer}$ is the average between that of
dry PSS (1.53 \cite{Theo_thesis}) and that of water (1.33)
weighted by the respective volume fractions ($\Phi_{\rm
pearl}=\pi/(3\sqrt{3})\simeq 0.60$ for a compact hexagonal
monolayer) and one finds also a value of 1.45. Therefore, in what
follows, the layer index has been fixed to $n_{layer}=1.45$ and
$\Psi$ and $\Delta$ interpreted in terms of the apparent thickness
$h_{app}$.

\begin{figure}
\onefigure[scale=0.5]{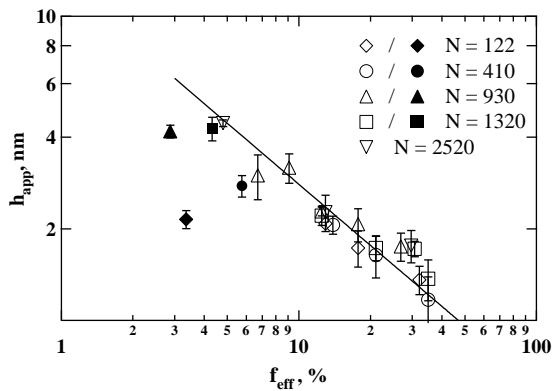}\caption{Apparent thickness
$h_{app}$ of adsorbed layers as a function of the effective charge
fraction $f_{eff}$ for PSS of various chain lengths. Each point is
an average of 12 measurements. For this adsorbing conditions
($C_{p}=0.01$ mol/L, [NaCl]=0.1 mol/L), this thickness can be
interpreted as the pearl size of PSS. The straight line has a
slope of -2/3. Filled points are reported in fig.~\ref{thickDP}. }
\label{thickness}
\end{figure}

The dependence of $h_{app}$ on the effective charge fraction
$f_{eff}$ is presented in fig.~\ref{thickness} for various chain
lengths $N$.
It shows clearly that $h_{app}$ depends significantly on the
effective charge fraction whereas it is independent of $N$ (if one
ignores the filled symbols which will be considered later). Within
experimental errors $h_{app}$ obeys the following power law:
\begin{equation}
h_{app}\sim aN^0f_{eff}^{-2/3} \label{happ1}
\end{equation}
In order to interpret this charge density dependence we assume
that $h_{app}$ is closely related to the pearl size $D_p$ (PSS
layer viewed as a carpet of pearls as noted above). The $N$ and
$f$ exponents are indeed in remarkable agreement with that for
$D_p$ predicted for the bulk conformation (eq.~\ref{pearl
scaling}) in the pearl-necklace model. This confirms that we are
in the conditions where the pearl-necklace conformation persists
upon electrostatic adsorption as it has been predicted
theoretically in the case of highly charged hydrophobic
polyelectrolytes~\cite{Dobrynin_2002}. Thus we identify the $N$
and $f$ exponents found here to those for $D_p$.

\begin{figure}
\onefigure[scale=0.5]{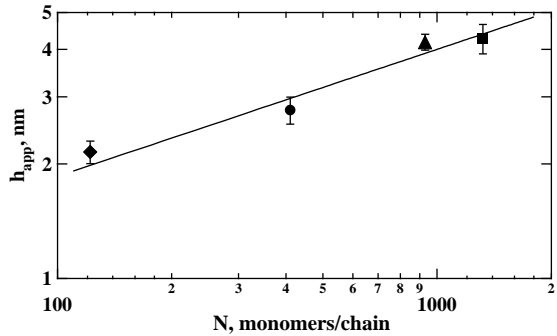} \caption{Apparent thickness
$h_{app}$ of adsorbed layers versus chain length $N$ for short PSS
($120\leq N\leq 1320$) at very low charge fractions
($f_{eff}<5\%$). The points correspond to filled points in
fig.~\ref{thickness}. The straight line has a slope of 1/3.}
\label{thickDP}
\end{figure}

Let us now come back to the filled points in fig.~\ref{thickness}
that deviate from the power law (eq.~\ref{happ1}) and correspond
to the lowest $N$ and $f$ values. They are reported in
fig.~\ref{thickDP} where $h_{app}$ for these points is plotted as
a function of the chain length.
Experimentally $h_{app}$ scales as
\begin{equation}
    h_{app}\sim aN^{1/3}
\end{equation}
Because in this case $f_{eff}$ is low the theoretically predicted
pearl size becomes very large and can even be larger than what the
number of monomers allows. This means than for these points (low
$N$, low $f_{eff}$), the chain is not envisioned as a
pearl-necklace with many pearls at equilibrium but instead as a
dense and spherical globule minimising its interfacial area with
the solvent. The globule size is less than $D_{p}$ and is stable
because its electrostatic energy is less than its interfacial
energy. Similarly to pearls, globules are expected to be close
compacted because of the very short-range repulsion. We thus
expect that $h_{app}$ is here proportional to the globule size.
The 1/3 exponent as represented by the straight line agrees indeed
with a compact globular structure of the collapsed chain. This is
an indirect evidence of a pearl-necklace/globule transition that
is controlled ar low $f$ by the chain length $N$.

It has to be pointed out that ellipsometry does not allow one to
characterise the shape of a single pearl or globule. Nevertheless
the identity of the scaling exponents indicate that only weak
deformations of the pearls (globules respectively) would take
place. This is to be expected with our experimental conditions of
strong screening of the electrostatic attraction to the surface
($\lambda_D < D_p$). Furthermore we believe that dipolar
attractive forces within the pearls (ionomer effect) stabilise the
conformation~\cite{note_ionomer}.

\begin{figure}
\onefigure[scale=0.5]{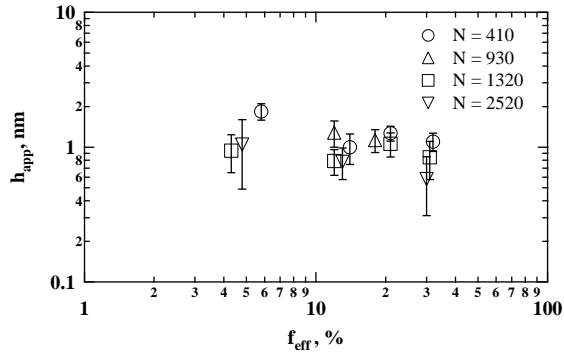} \caption{Apparent thickness
$h_{app}$ versus effective charge fraction $f_{eff}$ for PSS of
various chain lengths $N$ adsorbed from salt-free solutions. }
\label{thickSF}
\end{figure}

It is interesting to contrast these properties to that of layers
adsorbed from salt-free solutions. PSS of same chain lengths and
charge fractions were adsorbed from salt-free solutions at
$C_p=0.01$ mol/L. Because the free counter-ion concentration
($\simeq C_{p}f_{eff}$) is low, the Debye length $\lambda_{D}$ is
very large (10 to 30 nm) and one expects that the pearl-necklace
adopts a stretched conformation in the bulk. For a direct
comparison with the previous results the refractive index is
arbitrarily kept at $n_{\rm PSSlayer}=1.45$, the refractive index
of the dense carpet of pearls. The resulting apparent thickness
$h_{app}$ is plotted in fig.~\ref{thickSF} depending on $f_{eff}$
for PSS of various chain lengths. Within experimental errors
$h_{app}$ is found independent of both $f_{eff}$ and $N$. Using
the same $n_{layer}$ the film appears thin and the thickness is
smaller than the unperturbed pearl size. Two limit interpretations
can be envisioned. The first one assumes that the free chain
conformation remains in the adsorbed state. In this case pearls
still exist but they are separated by extended string immersed in
water. The apparent thickness appears constant because the string
length, the distance between two neighbouring pearls, increases
when the pearl size increases. The second one takes into account
the fact that the electrostatic attraction onto the oppositely
charged surface is much stronger than that in the salted regime
($\lambda_D \gg D_p$) and assumes that pearls spread on the
surface. A smooth monolayer is formed with a thickness of the
order of the thermal blob size $\xi_{T}$ which does not depend on
$f$ similarly to what has predicted a recent theoretical
model~\cite{Borisov_2001}. \emph{In situ} ellipsometry is not
capable to differentiate these two scenarii.

\section{Conclusion}
\begin{figure}
\onefigure[scale=0.5]{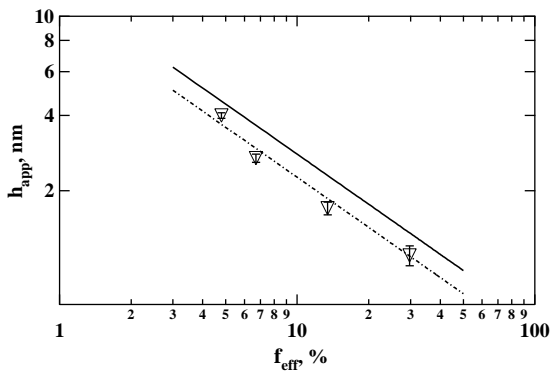} \caption{Apparent thickness
$h_{app}$ as a function of the effective charge fraction $f_{eff}$
as measured by high energy \emph{in situ} X-ray reflectivity. The
continuous line is that of fig.~\ref{thickness}. The dashed line
has also a slope of -2/3. Measurements were performed in a custom
made liquid cell using high energy (19~keV) X-ray from the ESRF
synchrotron (Grenoble, France) and will be detailed in a
forthcoming paper \cite{Baigl_refX}.} \label{thickrefX}
\end{figure}

A model hydrophobic polyelectrolyte, partially sulfonated
polystyrenesulfonate (PSS) of various chain lengths $N$ and
effective charge fractions $f_{eff}$, has been adsorbed onto
oppositely charged surfaces in conditions where the pearl-necklace
conformation predicted for the single chain in the bulk is
expected to be maintained in the adsorbed state. The apparent
layer thickness $h_{app}$ has been measured by \emph{in situ}
spectroscopic ellipsometry as a function of $f_{eff}$ and $N$ and
we have found $h_{app}\sim aN^0f_{eff}^{-2/3}$. If we assume that
$h_{app}$ is proportional to the pearl size, this would be the
first experimental confirmation of the predicted scaling law
(eq.~\ref{pearl scaling}) relating the pearl size to the charge
fraction in the pearl-necklace model~\cite{Dobrynin_1996}. At low
$f_{eff}$ we have found $h_{app}\sim aN^{1/3}$ for the shortest
chains. We interpret this power law as an evidence of a
globule/necklace transition at low $f$ controlled by $N$. Moreover
it is interesting to note that we have preliminary results from
\emph{in situ} high energy X-ray reflectivity measurements
\cite{Baigl_refX} that confirm the scaling law for $h_{app}$.
Fig.~\ref{thickrefX} displays the PSS thickness obtained by this
technique which indeed depends on the effective charge fraction
with the same -2/3 scaling exponent. On the other hand an
important feature is that the scaling laws reported here have been
expressed in terms of $f_{eff}$ rather than $f$, the bare charge.
This shows that $f_{eff}$ is the experimental parameter to take
into account for a correct estimation of the electrostatic
contribution. Further experiments will explore the influence of
the surface charge density and \emph{in situ} AFM in soft contact
mode will allow us to image the adsorbed PSS molecules.

\acknowledgments We wish to acknowledge J. L. Keddie and C.
Carelli (University of Surrey) for their helpful support during
the \emph{in situ} ellipsometry measurements.

\end{document}